\documentclass[12pt,a4paper]{article}
\usepackage{natbib}
\usepackage{soul}
\usepackage{latexsym}
\usepackage{amsmath,amsthm,amssymb}
\usepackage{rotating}
\usepackage{color}

\newcommand{\eins}{{\bf 1}}
\newcommand{\dif}{{\,\rm d}}

\renewcommand{\P}{\ensuremath{\mathrm{P}\,}}
\newcommand{\refeq}[1]{(\ref{#1})}

\input epsf.sty
\def\Prodi{\mathop{{\lower9pt\hbox{\epsfxsize=15pt\epsfbox{pi.ps}}
    }}}
\def\prodi{\mathop{{\lower3pt\hbox{\epsfxsize=7pt\epsfbox{pi.ps}}
}}}

\date{}
\begin{document}

\title{A competing risks approach for nonparametric estimation of
transition probabilities in a non-Markov illness-death model}
 \author{Arthur Allignol, Jan Beyersmann,  Thomas Gerds and Aur{\'e}lien Latouche}
   \date{\today}
\maketitle

\section*{} \textbf{Abstract}\hspace{0.5cm} Competing risks model time
to first event and type of first event. An example from hospital
epidemiology is the incidence of hospital-acquired infection, which
has to account for hospital discharge of non-infected patients as a
competing risk. An illness-death model would allow to further study
hospital outcomes of infected patients. Such a model typically relies
on a Markov assumption. However, it is conceivable that the future
course of an infected patient does not only depend on the time since
hospital admission and current infection status but also
on the time since infection. We demonstrate how a modified competing
risks model can be used for nonparametric estimation of transition
probabilities when the Markov assumption is violated. \vspace*{0.25cm}

\noindent {\textbf{Keywords:} Left-truncation $\cdot$
  Bivariate survival $\cdot$ Nosocomial Infection $\cdot$ Markov
  assumption $\cdot$ Multi-state model}\vspace*{0.5cm}

\section{Introduction}\label{sec:intro}
A competing risks model considers time to first event and type of
first event. In real life, one competing event, say event~1, may be
intermediate, and it could be of interest to investigate subsequent
occurrence of event~2. This is feasible by extending the competing
risks model to an illness-death model. The idea is that
  all individuals are initially subject to the original competing
  risks experiment. For those individuals who had a type~$1$ event as
  a first event, a second experiment determines the waiting time
  between the type~$1$ event and the type~$2$ event. See
\cite{fine2001semi} for a related extension of competing risks.

Both competing risks and illness-death models are, for instance,
relevant in hospital epidemiology \citep{jb:fogr:2011}: Nosocomial,
i.e., hospital-acquired infections are a major healthcare concern,
increasing morbidity and mortality, and they are a problem from a
health economics perspective.  \cite{umscheid2011estimating}
considered preventable nosocomial infections and argued that
successful prevention could save up to 53,483 lives a year in the
U.S., with up to \$23.44 billion annual cost savings to hospitals.

\cite{grambauer10:_incid} recently demonstrated that estimating the
incidence of nosocomial infections must account for end of hospital
stay without prior infection as a competing risk, i.e.,
  direct discharge of a patient prevents in-hospital infection.
Predicting length of hospital stay for an infected patient or
predicting the proportion of infected in-hospital patients is relevant
for the planning of hospital resources, but must account for the
time-dependency of the infection status as in an illness-death model
\citep{graves2011importance}. In this model, all patients
  would share one initial state. Infected patients move into the
  intermediate illness state at the time of infection, and end of stay
  is modelled by transitions into the absorbing state.

The canonical nonparametric estimator of the transition probabilities
in these models is the Aalen-Johansen estimator
\citep{Aale:Joha:an:1978}. The estimator relies on a
time-inhomogeneous Markov assumption, which is trivially fulfilled for
competing risks, but may be violated in an illness-death model. In the
context of nosocomial infections, the assumption does not hold, if the
end-of-hospital stay probability of an infected patient depends on the
time of infection.

{\color{black}Research for possibly non-Markov models has mostly
  focused on estimating state occupation probabilities~$P(X_t=j)$,
  where $X_t$ denotes the state occupied at time~$t$ and $j$ is a
  possible state of the model. Under a Markov assumption and assuming
  one initial state occupied by all individuals at time~$0$,
  say~$P(X_0=0)=1$, estimation may be based on the Aalen-Johansen
  estimator of~$P(X_t=j\,|\, X_0=0)$. In the absence of a common
  initial state, the Aalen-Johansen estimator of~$P(X_t=j\,|\, X_0=\cdot)$
  would need to be multiplied by an estimator of the initial state
  distribution.

  For complete data, \cite{ABGK} showed that this approach equals the
  usual multinomial estimators which do not rely on a Markov
  assumption. A major breakthrough for data subject to random
  right-censorship was then obtained by \cite{Datt:Satt:vali:2001} and
  \cite{Glid:robu:2002}. Datta and Satten showed that this
  Aalen-Johansen approach still consistently estimates the state
  occupation probabilities in the absence of the Markov property, and
  Glidden provided weak convergence results. Earlier work of
  \cite{pepe1991qualifier} had allowed for estimating the probability
  of an intermediate condition in a non-Markov illness-death model.
  Interestingly, Pepe et al.\ found their estimator to approximately
  equal the standard Aalen-Johansen estimator, somewhat anticipating
  the subsequent more general results of Datta and Satten.

  \cite{Datt:Satt:esti:2002} allowed for non-random censoring by
  directly modelling the censoring hazard; see also related results by
  \cite{Datt:Satt:Datt:nonp:2000} for the illness-death model.
  \cite{gunnes2007estimating} discussed the relative merits of the
  Aalen-Johansen and the Datta-Satten estimator in terms of bias and
  mean squared error in the presence of dependent censoring. See
  \cite{datta12:_recen_advan_system_reliab} for an overview.

  A different line of research that could be applied to non-Markov
  multistate models is time-multivariate survival analysis.
  \cite{Gill:mult:1992} mentions this possibility and gives an
  insightful discussion on why nonparametric estimation of a
  multivariate survival function in the presence of multivariate
  censoring is a difficult problem, where the usual counting process
  approach breaks down. \cite{Lin:Ying:a-si:1993} noted that the
  difficulties reduce and simpler estimation procedures are feasible,
  if censoring is univariate. This is the case in a multistate model.
  \cite{tsai} improved on the Lin-Ying estimator, and an overview was
  given by \cite{prentice04:_nonpar}.  }

The aim of the present paper is to use competing risks techniques for
nonparametric estimation of transition probabilities in a potentially
non-Markov illness-death model without recovery. {\color{black} This
  aim differs from estimating state occupation
  probabilities~$P(X_t=j)$ in that we do wish to condition on the
  state occupied at time~$s$, $s\le t$. There is a connection to
  time-multivariate survival analysis, because the first estimator
  that we will derive is algebraically identical to an earlier
  proposal by \cite{Meir:Ua:Cada:nonp:2006}. To the best of our
  knowledge, the work by Meira-Machado et al.\ was the first paper
  which focused on using time-multivariate techniques for estimation of
  transition probabilities in a non-Markov illness-death model,
  employing the time-multivariate techniques of
  \cite{stute93:_consis}.

  We develop the Meira-Machado et al.\ estimator via a different
  route, which allows for a competing risks explanation on why their
  estimator works in a non-Markov model. We also give a new inverse
  probability of censoring weighted (IPCW) representation of the
  estimator. Using both the new IPCW representation and results of
  \cite{tsai}, we derive a new, simpler and theoretically more
  efficient competing risks-type estimator. The new estimator gives
  direct access to competing risks methodology, which we demonstrate
  by also allowing for left-truncation.}

The paper is organized as follows: Section~\ref{sec:models} introduces
competing risks and illness-death models as stochastic processes. The
illness-death model is also re-parametrized via a bivariate time
vector and a further competing risks model is derived, which will be
crucial for the nonparametric estimation procedures of
Section~\ref{sec:est}. We report simulation results in
Section~\ref{sec:simu} and an analysis of real hospital infection data
in Section~\ref{sec:illu}. The closing Section~\ref{sec:disc} offers a
discussion, including an appraisal of the relative merits of the 
Meira-Machado et al.\ estimator and the new competing risks estimator.
Our conclusion is that both estimators perform comparably, but that
the new estimator may be preferred due to its computational
simplicity. {\color{black}We also find that the Aalen-Johansen
  estimator may perform competitively even if the Markov assumption is
  violated.}

\section{Competing risks and illness-death models}\label{sec:models}
Consider a stochastic process~$(X_u)_{u\in[0,\infty)}$ with state
space~$\{0,1,2\}$, right-continuous sample paths and initial
state~$0$, $P(X_0=0)=1$.
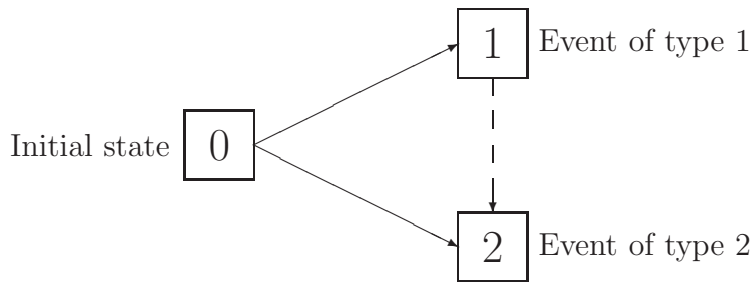
\begin{figure}[htb]
  \centering
  \setlength{\unitlength}{0.9cm}
  \begin{picture}(5,5)(0.6,-2)
    
    \put(0,0){\framebox(1,1){\Large 0}}
    \put(0,0){\makebox(1,1){\parbox{5.5cm}{Initial state}}}
    
    \put(1,0.5){\vector(2,1){3}} \put(4,1.5){\framebox(1,1){\Large 1}}
    \put(1,0.5){\vector(2,-1){3}}\put(4,-1.5){\framebox(1,1){\Large 2}}

    \put(4.5,1.475){\line(0,-1){0.25}}\put(4.5,1){\line(0,-1){0.25}}
    \put(4.5,0.5){\line(0,-1){0.25}}\put(4.5,0){\vector(0,-1){0.5}}
    
    \put(8,1.5){\makebox(1,1){\parbox{6cm}{Event of type $1$}}}
    \put(8,-1.5){\makebox(1,1){\parbox{6cm}{Event of type $2$}}}

  \end{picture}      
  \caption{Competing risks model and illness-death model without recovery.}
  \label{fig:fig}
\end{figure}
For a competing risks model with two competing risks, we model~$0\to
1$ and~$0\to 2$ transitions, and states~$1$ and~$2$ are absorbing,
i.e., there are no transitions out of the absorbing states. In the
context of nosocomial infections, we will consider patients to enter
state~$0$ on admission to hospital. Occurrence of an infection is
modelled by a $0\to 1$ transition, end of hospital stay without prior
infection is modelled by a $0\to 2$ transition.

We may extend this model to an illness-death model without recovery by
also allowing for $1\to 2$ transitions. This is illustrated in
Figure~\ref{fig:fig}, where the dashed arrow indicates that $1\to 2$
transitions are only feasible in the illness-death model. The addendum
`without recovery' means that $1\to 0$ transitions are not modelled.
For nosocomial infections, this entails that $X_u=1$ is interpreted as
`in hospital at time~$u$, infection has occurred in~$(0,u]$'. This
interpretation is in line with the common comparison of infected
`cases' and non-infected `controls' in hospital
epidemiology. {\color{black} The interpretation of $X_u=2$ is that
  hospital stay has ended by time~$u$.}

Also note that the interpretation of states~$1$ and~$2$ differs
between the models. For competing risks, the interpretation of
state~$1$ is `an infection has occurred', while the interpretation of
state~$2$ is `hospital stay has ended without prior infection'.

Regardless of the model, we may define the time until first event,
\begin{equation}
  \label{eq:T0}
  T_0 = \inf\{u\;:\; X_u\neq 0\}.
\end{equation}
The type of first event is
\begin{equation}
  \label{eq:XT0}
  X_{T_0}\in\{1,2\},
\end{equation}
the state entered by the process at time~$T_0$.  

For the illness-death model, we also define the time until absorption
(end of hospital stay),
\begin{equation}
  \label{eq:T}
  T = \inf\{u\;:\; X_u=2\}.
\end{equation}
We have $T_0=T$, if the process makes a direct $0\to 2$ transition,
and $T_0<T$ otherwise. We assume that the distribution of $T$ has mass
on~$[0,\infty)$ only. That is, every individual reaches state~$2$
(spends a finite time in hospital).

In the remainder of the paper, we will take $(X_u)_u$ to be an
illness-death model. The aim will be to provide for non-parametric
estimation of the transition probabilities
\begin{equation}
  \label{eq:P}
  P_{lj}(s,t) = P(X_t=j\,|\,X_s=l),
\end{equation}
where $(s,t)$, $s\le t$, is a fixed, but arbitrary pair of times,
$l\in\{0,1\}$, $j\in\{1,2\}$. In \eqref{eq:P}, we do not assume that
conditioning on $X_s=1$ is tantamount to conditioning on the entire
past of the process up to time~$s$. That is, we do not assume that
$(X_u)_u$ is Markov.

More specifically and for ease of presentation, we will focus on
$P_{01}(s,t)$. In the data example, this is the probability of an
infected in-hospital patient at time~$t$ given no infection at
time~$s$. This quantity can be used for the planning of hospital
resources. Our ideas work analogously for the other transition
probabilities. We express $P_{01}(s,t)$ in terms of the bivariate time
vector~$(T_0,T)$,
\begin{equation}
  \label{eq:2}
  P_{01}(s,t) = \frac{P(X_t=1,X_s=0)}{P(X_s=0)} = \frac{P(s<T_0\le t, t<T)}{P(T_0>s)}.
\end{equation}

The key to the nonparametric estimation procedures in
Section~\ref{sec:est} are both~\eqref{eq:2} and the following
competing risks
process~$(\kappa_{u;s,t})_u=(\kappa_u)_u$, which is
derived from~the illness-death process $(X_u)_u$,
\begin{equation}
  \label{eq:kappa}
  \kappa_{u;s,t} = \kappa_u = \left\{
    \begin{array}{r@{\quad:\quad}l}
      0 & X_u \in\{0,1\},\\
      1 & X_u=2\mbox{\ and\ }\eins(s<T_0\le t, t<T)=1,\\
      2 & X_u=2\mbox{\ and\ }\eins(s<T_0\le t, t<T)=0,
    \end{array}\right.
\end{equation}
where~$\eins(\cdot)$ is the indicator function. The competing risks
process~$\kappa$ stays in its initial state~$0$ until time~$T$. At
time~$T$, the value of the competing risks mark~$\eins(s<T_0\le t,
t<T)$ is known. We have that~$P(\kappa_T=1)=P(s<T_0\le t, t<T)$. As a
consequence, the numerator of the right hand side of~\eqref{eq:2} is
the limit of the cumulative incidence function for event
type~$1$ of~$\kappa$,
\begin{equation}
  \label{eq:limit}
  P(s<T_0\le t, t<T) = \lim_{u\to\infty} P(T\le u, \kappa_T=1).
\end{equation}
Note that the competing risks process~$\kappa$ depends on
  the fixed, but arbitrary pair of times $(s,t)$, $s\le t$, but we are
  suppressing this in the notation for ease of writing.

\section{Nonparametric estimation}\label{sec:est}
We assume that observation of the illness-death process~$X$, or,
equivalently, of the random times~$(T_0,T)$, is subject to random
censorship by~$C$. We also assume that the support of the distribution
of~$T$ is contained in the support of the distribution of~$C$. This
last assumption is needed for estimation of the limit of the
cumulative incidence function in~\eqref{eq:limit}. It is justifiable
for the nosocomial infection example, but may be violated in other
settings. In the discussion, we explain how this assumption can be
relaxed. 
{\color{black} We first revisit the estimator of
  \cite{Meir:Ua:Cada:nonp:2006} in Section~\ref{sec:est:sub1},
  revealing that violations of the Markov assumption can be seen to be
  handled via a competing risks approach and also giving a new IPCW
  representation of the estimator. These two observations are taken
  further in Section~\ref{sec:est:sub2}, leading to a simpler
  competing risks-type estimator, which in turn also allows for
  left-truncated data as explained in Section~\ref{sec:est:sub3}.}

{\color{black}\subsection{The estimator of Meira-Machado et al.\
    revisited}}\label{sec:est:sub1} For estimation of~\eqref{eq:2}, we
use the usual Kaplan-Meier estimator for estimating the
denominator~$P(T_0>s)$, based on the censored observations of~$T_0$.
Because of~\eqref{eq:limit}, we use the right hand limit of the
Aalen-Johansen estimator of~$P(T\le u, \kappa_T=1)$ for estimation of
the numerator. To this end, and for the competing risks
process~$\kappa$, we write~$N_1$ for the counting process of observed
events of type~$1$, $N$ for the counting process of observed events
(of any type), and~$Y$ for the at-risk process. We also write $N_0$
for the counting process of observed replicates of~$T_0$ and $Y_0$ for
the at risk process of the initial state of the illness-death
model~$X$. {\color{black}Note that the processes $N_1$, $N$ and $Y$
  depend on the fixed pair of times $(s,t)$ through $\kappa$, but
  $N_0$ and $Y_0$ do not depend on $(s,t)$.} Then, these estimators
are
\begin{equation}
  \label{eq:est0}
  \hat{P}(T_0>s)=\Prodi_{v\in[0,s]}\left(1-\frac{\dif N_0(v)}{Y_0(v)}\right)
\end{equation}
and
\begin{equation}
  \label{eq:est1}
  \hat{P}(s<T_0\le t, t<T) = \int_0^\infty \Prodi_{v\in[0,u)}\left(1-\frac{\dif N(v)}{Y(v)}\right)\frac{\dif N_1(u)}{Y(u)}.
\end{equation}
{\color{black}Recall that the right hand side of~\refeq{eq:est1}
  depends on $(s,t)$ via $N_1$, $N$ and $Y$.}
{\color{black}In the appendix, we show} that the
resulting estimator of~$P_{01}(s,t)$,
\begin{equation}
  \label{eq:mm}
  \hat{P}_{01}(s,t) = \hat{P}(s<T_0\le t, t<T)/\hat{P}(T_0>s),
\end{equation}
equals the estimator proposed by \cite{Meir:Ua:Cada:nonp:2006}, who
derived their estimator via a different route, using Kaplan-Meier
integrals. Note that the estimator~\refeq{eq:mm} is, in
  general, different from the Aalen-Johansen estimator. This is even
  true for the simple case of $s=0$. Here, as a function of~$t$, the
  Aalen-Johansen estimator of~$P_{01}(0,t)$ will change its value
  whenever there is an observed $0\to 1$ transition in the
  illness-death model. In contrast, and assuming no ties, the
  non-Markov estimator will \emph{not} change its value (as a function
  of~$t$), if the individual at hand is subsequently censored in the
  intermediate state of the illness-death model. This is so, because
  $N_1$ is the counting process of observed events of type~$1$ of the
  competing risks process~$\kappa$. The event times of~$\kappa$ are
  the waiting times until absorption of the illness-death process.

We now {\color{black}give a new IPCW representation of the estimator,
  which we will subsequently use to} modify and thereby simplify
estimation of~$P_{01}(s,t)$. The idea is to express~\eqref{eq:est1} in
terms of a Kaplan-Meier estimator of the censoring survival function
and to {\color{black}then} use an observation by \cite{tsai}, who noted
that there is more than one such estimator in bivariate time.

We write $N^C$ for the counting process of censoring events, which
have been observed before absorption. We have that
\begin{displaymath}
  \Delta N^C(u) + \Delta N(u) + Y(u+) = Y(u),
\end{displaymath}
where~$\Delta$ indicates the increment of the respective processes. As
a consequence,
\begin{equation}
  \label{eq:censi}
  \Prodi_{v\in[0,u)}\left(1-\frac{\dif N(v)}{Y(v)}\right) \cdot
  \Prodi_{v\in[0,u)}\left(1-\frac{\dif N^C(v)}{Y(v)-\Delta
      N(v)}\right) = \frac{Y(u)}{Y(0)},
\end{equation}
and the estimator in \eqref{eq:est1} equals
\begin{equation}
  \label{eq:ipcw}
  \frac{1}{Y(0)}\int_0^\infty \Prodi_{v\in[0,u)}\left(1-\frac{\dif N^C(v)}{Y(v)-\Delta
      N(v)}\right)^{-1}\dif N_1(u).
\end{equation}
Here, $\prodi_{v\in[0,u)}\left(1-\frac{\dif N^C(v)}{Y(v)-\Delta
    N(v)}\right)$ is the Kaplan-Meier estimator of $P(C\ge u)$, based
on the censored observations of~$T$.
{\color{black}\subsection{A new competing risks-type estimator}\label{sec:est:sub2}} 
\cite{tsai} observed that there is more than one Kaplan-Meier-type
estimator of $P(C\ge u)$, if a bivariate vector of event times such
as~$(T_0, T)$ is subject to one censoring variable~$C$. We introduce
some additional notation: We write $N_0^C$ for the counting process of
censoring events, which have been observed before leaving the initial
state of the illness-death model~$X$. We also write ${}_sY$ for the at
risk process of the competing risks model~$\kappa$ in the data subset
of individuals who were still in the initial state of~$X$ and under
observation at time~$s$. We analogously define~${}_sN$, ${}_sN_1$ and
${}_sN^C$. Then Tsai and Crowley suggested to use the following
Kaplan-Meier-type estimator of $P(C\ge u)$, specialized to our setting
with $T_0\le T$,
\begin{equation}
  \label{eq:TC}
  \Prodi_{v\in[0,s]}\left(1-\frac{\dif N^C_0(v)}{Y_0(v)-\Delta
      N_0(v)}\right) \cdot \Prodi_{v\in(s,u)}\left(1-\frac{\dif {}_sN^C(v)}{{}_sY(v)-\Delta
      {}_sN(v)}\right).
\end{equation}
{\color{black} Replacing $\prodi_{v\in[0,u)}\left(1-\frac{\dif
      N^C(v)}{Y(v)-\Delta N(v)}\right)$ in~\eqref{eq:ipcw}
  by~\eqref{eq:TC} as an estimator of $P(C\ge u)$,}
we obtain a different estimator of $P(s<T_0\le t,
t<T)$,
\begin{eqnarray}
  \check{P}(s<T_0\le t, t<T) & = &\frac{1}{Y(0)} \Prodi_{v\in[0,s]}\left(1-\frac{\dif N^C_0(v)}{Y_0(v)-\Delta
      N_0(v)}\right)^{-1} \cdot\nonumber\\
  {} & {} & \int_0^\infty \Prodi_{v\in(s,u)}\left(1-\frac{\dif {}_sN^C(v)}{{}_sY(v)-\Delta
      {}_sN(v)}\right)^{-1} \dif N_1(u).\nonumber
\end{eqnarray}
Because~$Y(0)=Y_0(0)$, $Y_0(s+)={}_sY(s+)$ and (as a consequence of
the definition of~$\kappa$) $N_1={}_sN_1$, this equals
\begin{displaymath}
  \Prodi_{v\in[0,s]}\left(1-\frac{\dif N_0(v)}{Y_0(v)}\right) \int_s^\infty \Prodi_{v\in(s,u)}\left(1-\frac{\dif {}_sN(v)}{{}_sY(v)}\right)\frac{\dif {}_sN_1(u)}{{}_sY(u)}
\end{displaymath}
where we have also used an analogous variant of~\eqref{eq:censi}
for~$\hat{P}(T_0>s)=\prodi_{v\in[0,s]}\left(1-\frac{\dif
    N_0(v)}{Y_0(v)}\right)$.

The resulting estimator of~$P_{01}(s,t)$ is
\begin{eqnarray}
  \check{P}_{01}(s,t)& =& \check{P}(s<T_0\le t, t<T)/\hat{P}(T_0>s)\nonumber\\
  {} & = & \int_s^\infty \Prodi_{v\in(s,u)}\left(1-\frac{\dif {}_sN(v)}{{}_sY(v)}\right)\frac{\dif {}_sN_1(u)}{{}_sY(u)}.  \label{eq:check}
\end{eqnarray}
The estimator in~\eqref{eq:check} is simple: It is just an estimator
of the limit of a cumulative incidence function as in~\eqref{eq:est1},
but evaluated in the data subset `still in the initial state of~$X$
and under observation at time~$s$'.

{\color{black}Standard competing risks arguments can be used to derive
  an estimator of the variance of~$\check{P}_{01}(s,t)$ \citep[][p.~299]{ABGK}, 
  \begin{eqnarray*}
    \lefteqn{\hat{\rm var}\check{P}_{01}(s,t) =}\\
    {} & {} & \left.\right.\hspace{-0.75cm}\int_s^\infty \Bigg\{\Prodi_{v\in(s,u]}\left(1-\frac{\dif {}_sN(v)}{{}_sY(v)}\right)\Bigg\}^2\Bigg\{1- \\{} & {} & \int_u^\infty \Prodi_{v\in(u,r)}\left(1-\frac{\dif {}_sN(v)}{{}_sY(v)}\right)\frac{\dif {}_sN_1(r)}{{}_sY(r)}\Bigg\}^2 \frac{\dif {}_sN_1(u)}{{}_sY(u)} \quad\quad + \\
    {} & {} & \left.\right.\hspace{-0.75cm}\int_s^\infty \Bigg\{\Prodi_{v\in(s,u]}\left(1-\frac{\dif {}_sN(v)}{{}_sY(v)}\right)\cdot\int_u^\infty \Prodi_{v\in(u,r)}\left(1-\frac{\dif {}_sN(v)}{{}_sY(v)}\right)\frac{\dif {}_sN_1(r)}{{}_sY(r)}\Bigg\}^2 \frac{\dif {}_sN_2(u)}{{}_sY(u)},
  \end{eqnarray*}
  where we have also used ${}_sN_2$ for the counting process of
  observed events of type~$2$ of the competing risks model~$\kappa$ in
  the data subset of individuals who were still in the initial state
  of~$X$ and under observation at time~$s$.
  {\color{black}
  This variance estimator is motivated by a corresponding asymptotic expression \citep[p.~321.]{ABGK}.
   
  Theoretically, the new estimator is more efficient than the one of
  \citet[Theorem 2]{Meir:Ua:Cada:nonp:2006}. The informal argument is that 
  it uses the full information from the subjects whose
  illness-death process was right censored, whereas the Meira-Machado et al.\ 
  estimator ignores the information in which state the subjects were right censored. 
  This can be seen by comparing the weights used in the construction of the IPCW estimators
  (this is were information from the censored subjects enters).
  The new estimator uses the conditional weights given in \refeq{eq:TC}.}
 The first factor of~\refeq{eq:TC} estimates $P(C\ge s \,|\, T_0>s)$ using all
  censored times that are less than or equal to time~$s$ and where the
  corresponding illness-death process is \textit{censored in the
    initial state}. The second factor estimates $P(C\ge u \,|\, T_0>s,
  C>s)$ using all the censoring times that are greater than time $s$
  and less than or equal to time~$u$ for which the corresponding
  illness-death process is \textit{in the initial state and under
    observation at time~$s$.}
  {\color{black}
  The Meira-Machado et al.\ estimator uses IPCW weights derived from the
  marginal Kaplan-Meier estimator $P(C\ge u)$  which uses the censoring times but not the
  state of the illness-death process at the individual censoring time.
  There are similar results and a general theory for IPCW \citep{van:Robi:unif:2003}
  which could be used to show formally that $\check{P}_{01}(s,t)$ is
  asymptotically more efficient as compared to $\hat{P}_{01}(s,t)$.
  However, our simulation results and data example show comparable
  small sample performances of both estimates (see Sections \ref{sec:simu} and \ref{sec:illu}).}

  \subsection{Left-truncated data}\label{sec:est:sub3}
  
  So far, we have assumed that observation of the illness-death
  process is subject to random censoring only. We now additionally
  allow for left-truncation (delayed study entry), which can be
  handled by the new estimator $\check{P}_{01}(s,t)$ because of
  general competing risks results \citep{ABGK}. To be specific, assume
  that observation of the random times~$(T_0,T)$, is subject to random
  left-truncation and right-censorship by~$(L,C)$, i.e., we assume
  that the tuples~$(T_0,T)$ and~$(L,C)$ are independent.

  We have to be precise what delayed study entry in this context
  means, because $\check{P}_{01}(s,t)$ is an estimated cumulative
  incidence function, estimated in the data subset `in the initial
  state of~$X$ and under observation at time~$s$'. This entails that
  only an individual whose left-truncation time~$L$ is less than its
  waiting time~$T_0$ in the initial state can enter the calculation.
  This is in contrast to standard nonparametric estimation for a
  time-inhomogeneous Markov model, where an individual may be in any
  non-absorbing state of the model at the time of study entry.

  We now write ${}_sY$ for the at risk process of the competing risks
  model~$\kappa$ in the data subset of individuals \emph{whose
    left-truncation times were less than s and} who were still in the
  initial state of~$X$ and under observation at time~$s$. We
  analogously interpret~${}_sN$, ${}_sN_1$ and ${}_sN_2$. We can then
  profit from the general fact that counting processes naturally
  account for left-truncation \citep{keiding1992independent} and
  estimate~$P_{01}(s,t)$ using
  \begin{displaymath}
    \check{P}_{01}(s,t) = \int_s^\infty \Prodi_{v\in(s,u)}\left(1-\frac{\dif {}_sN(v)}{{}_sY(v)}\right)\frac{\dif {}_sN_1(u)}{{}_sY(u)}.
  \end{displaymath}
  At the beginning of the section, we had been forced to assume the
  support of the distribution of~$T$ to be contained in that of~$C$,
  because integrals as on the right hand side of the previous display
  are being evaluated up to~$\infty$. We now need to additionally
  account for the presence of left-truncation. Essentially what we
  need to ensure is that the risk set~${}_sY$ is non-empty
  on~$[s,\infty)$ with asymptotic probability larger than zero. To be
  precise, we assume that for all~$u < \inf\{v:\P(T>v)=0\}$ there
  exists a positive function~$y$ on~$[0,u]$, bounded away from zero,
  such that
  \begin{displaymath}
    \sup_{v\in [s,u]} \left|{}_sY(v)/{}_sY(s+) - y(v)\right|\to 0
  \end{displaymath}
  in probability as the `sample size'~${}_sY(s+)$ goes to infinity
  \citep[][Condition~(4.1.16)]{ABGK}.

}

\section{Simulation Study}\label{sec:simu}
We now report results of a limited simulation study, where the aim is
to compare the finite sample performance of our new
estimator~$\check{P}_{01}(s,t)$ from~\eqref{eq:check} with the more
complicated estimator~$\hat{P}_{01}(s,t)$ from~\eqref{eq:mm}, which is
algebraically equal to the estimator of \cite{Meir:Ua:Cada:nonp:2006}.
{\color{black}We also report results from using the Aalen-Johansen
  estimator.}

We simulated data from a scenario used by Meira-Machado et al., which
these authors found to be challenging both in terms of bias and
variance. To be specific, we generated replicates of $(T_0,X_{T_0})$
using an exponential hazard of $0.039 + 0.026$ for simulating $T_0$
and deciding on $X_{T_0}=1$ in a binomial experiment with
probability~$0.039/(0.039 + 0.026)$. If $X_{T_0}=1$, we set $T=
1.7\cdot T_0$; as a consequence, the model is not Markov. Random
censoring was simulated from an exponential distribution with
parameters $0.013$ or $0.035$. {\color{black}In addition, we also
  investigated $\check{P}_{01}(s,t)$ when the data were subject to
  both left-truncation and right-censoring. Left-truncation was
  simulated from a skew normal distribution \citep{skewnormal}, with
  location equal to $-5$, scale equal to 10 and shape equal to 10.
  Right-censoring was exponentially distributed with hazard~$0.013$.}

We simulated~$1000$ studies and report the bias (average of the $1000$
estimates of $P_{01}(10,t)$ minus true quantity) and the empirical
variance of the estimates. {\color{black}In the presence of
  right-censoring only, the sample size in each simulated study
  was~100. With additional left-truncation, the average sample size
  was~$85$.} The true value $P_{01}(10,t)$ was numerically
approximated based on $100$ replications of uncensored samples of size
$10000$ using the usual binomial estimator within the data subset
defined by `in state~$0$ at time~$10$', {\color{black}yielding
  
  \begin{tabular}{l|rrrrrrrrr}
    $t$ & 30 & 40 & 50 & 60 & 70 & 80 & 90 & 100\\\hline
    $P_{01}(10,t)$ & 0.201 & 0.162 & 0.125 & 0.092 & 0.067 & 0.048 & 0.033 & 0.023
  \end{tabular}
}

Tables~\ref{tab1} and~\ref{tab2} give results {\color{black}for the
  right-censoring scenarios, table~\ref{tab3} displays results for the
  scenario subject to both left-truncation and right-censoring.}

The tables indicate similar performance of both
estimators~\eqref{eq:mm} and~\eqref{eq:check} in terms of bias and
variance {\color{black}and in the presence of right-censoring only}.
Similar results were found for a sample size of~$200$ (not shown).
{\color{black}Interestingly, Tables~\ref{tab1} and~\ref{tab2} find the
  Aalen-Johansen estimator to perform at least competitively except
  for the early time point~$30$. This is somewhat in contrast to the
  results reported by Meira-Machado et al., who found the
  Aalen-Johansen estimator to be biased in the absence of the Markov
  property. The reason is that these authors considered the absolute
  bias integrated over time, which appears to be dominated by early
  time points. We find a similar picture when comparing the new
  estimator and the Aalen-Johansen in the presence of additional
  left-truncation.}

\begin{table}
\begin{center}

\begin{tabular}{rrrrrrrrr}
\hline
  &  \multicolumn{2}{c}{$\check{P}_{01}(10,t)$} & \multicolumn{2}{c}{$\hat{P}_{01}(10,t)$} &  \multicolumn{2}{c}{\color{black} Aalen-Johansen}   \\
$t$  &  Bias &Variance &  Bias &Variance &  Bias &Variance \\ 
  \hline
30 & 1.92e-03 & 5.07e-03 & 1.91e-03 & 5.02e-03 & -2.10e-02 & 3.92e-03 \\ 
  40 & 4.69e-03 & 4.46e-03 & 4.74e-03 & 4.45e-03 & -7.44e-03 & 3.57e-03 \\ 
  50 & -3.33e-03 & 4.44e-03 & -3.21e-03 & 4.46e-03 & -5.75e-03 & 3.62e-03 \\ 
  60 & -6.42e-03 & 3.86e-03 & -6.35e-03 & 3.88e-03 & -3.14e-03 & 3.08e-03 \\ 
  70 & -1.05e-02 & 3.05e-03 & -1.05e-02 & 3.06e-03 & -2.90e-03 & 2.54e-03 \\ 
  80 & -8.47e-03 & 2.39e-03 & -8.49e-03 & 2.39e-03 & 1.26e-03 & 2.17e-03 \\ 
  90 & -9.61e-03 & 1.51e-03 & -9.62e-03 & 1.51e-03 & 1.71e-03 & 1.60e-03 \\ 
  100 & -7.02e-03 & 1.11e-03 & -7.03e-03 & 1.11e-03 & 5.05e-03 & 1.37e-03 \\ 
   \hline
\end{tabular}
\end{center}\caption{Simulation results for censoring hazard~$0.013$.}\label{tab1}
\end{table}
%
%
\begin{table}
\begin{center}
\begin{tabular}{rrrrrrr}
\hline
 &  \multicolumn{2}{c}{$\check{P}_{01}(10,t)$} & \multicolumn{2}{c}{$\hat{P}_{01}(10,t)$} &  \multicolumn{2}{c}{\color{black} Aalen-Johansen}   \\
 
$t$  &  Bias &Variance &  Bias &Variance &  Bias &Variance \\ 
 \hline
30 & 3.31e-03 & 1.28e-02 & 2.92e-03 & 1.27e-02 & -1.61e-02 & 7.27e-03 \\ 
  40 & -1.14e-02 & 1.54e-02 & -1.16e-02 & 1.53e-02 & -4.94e-03 & 9.52e-03 \\ 
  50 & -3.35e-02 & 1.29e-02 & -3.36e-02 & 1.28e-02 & -6.03e-03 & 9.48e-03 \\ 
  60 & -3.78e-02 & 8.93e-03 & -3.80e-02 & 8.82e-03 & 3.41e-03 & 8.86e-03 \\ 
  70 & -4.14e-02 & 4.89e-03 & -4.15e-02 & 4.87e-03 & 9.20e-03 & 8.55e-03 \\ 
  80 & -3.39e-02 & 2.78e-03 & -3.39e-02 & 2.75e-03 & 2.04e-02 & 8.36e-03 \\ 
  90 & -2.75e-02 & 1.03e-03 & -2.76e-02 & 1.01e-03 & 2.82e-02 & 7.74e-03 \\ 
  100 & -2.08e-02 & 3.94e-04 & -2.08e-02 & 3.86e-04 & 3.56e-02 & 7.58e-03 \\ 
   \hline
\end{tabular}
\end{center}\caption{Simulation results for censoring hazard~$0.035$.}\label{tab2}
\end{table}

\begin{table}\color{black}
\begin{center}
\begin{tabular}{rrrrr}
 
  &  \multicolumn{2}{c}{Aalen-Johansen} &  \multicolumn{2}{c}{$\check{P}_{01}(10,t)$} \\
$t$ & Bias& Variance & Bias & Variance \\ 
  \hline
     30 & -2.17e-02 & 4.00e-03 & 3.18e-04 & 5.41e-03 \\ 
  40 & -9.38e-03 & 4.03e-03 & 2.06e-03 & 5.24e-03 \\ 
  50 & -5.30e-03 & 3.55e-03 & -1.33e-03 & 4.62e-03 \\ 
  60 & -1.38e-03 & 3.05e-03 & -2.79e-03 & 4.02e-03 \\ 
  70 & -4.83e-04 & 2.42e-03 & -6.90e-03 & 3.02e-03 \\ 
  80 & 1.25e-03 & 2.02e-03 & -8.43e-03 & 2.28e-03 \\ 
  90 & 2.38e-03 & 1.69e-03 & -9.27e-03 & 1.59e-03 \\ 
  100 & 3.85e-03 & 1.38e-03 & -9.10e-03 & 9.97e-04 \\ 
    \hline
\end{tabular}
 \end{center}\caption{Simulation results for left truncated data and  censoring hazard~$0.013$}\label{tab3}
\end{table}

\section{Real data example}\label{sec:illu}
We use a random subsample of 1313 patients from the SIR3
(\emph{S}pread of nosocomial \emph{I}nfections and \emph{R}esistant
pathogens) study that has been made publicly available as part of the
R-package {\bf kmi} \citep{bam}. The present analyses may therefore be
reproduced. SIR3 was a prospective study to assess the occurrence and
the impact of hospital-acquired infections in intensive care. Details
are reported elsewhere
\citep{Beye:Gast:Grun:Brwo:Geff:Behn:Rden:Schu:asse:2005}. Here, we
focus on the occurrence of hospital-acquired pneumonia, which is one
of the most frequent and most severe nosocomial infections. In an
analysis of the full data set of~$1876$ patients,
\cite{alli:cost:2010} included time of pneumonia as a time-dependent
covariate into Cox models for the end-of-stay hazards (distinguishing
between competing endpoints alive discharge and hospital death).
Because the hazard ratios were approximately equal to one in this
informal check of the Markov assumption, these authors concluded that
one may assume the data to follow a time-inhomogeneous Markov model.
However, because the confidence intervals were marginal, a more robust
estimation procedure as in the present paper may be desirable.

Tables~\ref{tab:s3},~\ref{tab:s5} and~\ref{tab:s7} report results on
estimating~$P_{01}(s,t)$ for $s = 3$, $s = 5$ and $s = 7$, using both
$\hat{P}_{01}(s,t)$ and $\check{P}_{01}(s,t)$. These estimates are
relevant for planning hospital resources, estimating the probability
of future infected intensive care patients among the currently, i.e.,
at time~$s$ uninfected.

The tables also report variance estimates and 95\% confidence
intervals (CI) computed from 1000 bootstrap samples. We used the
bootstrap in order to have one common method for both
$\hat{P}_{01}(s,t)$ and $\check{P}_{01}(s,t)$. Section~\ref{sec:est}
has shown that estimating a cumulative incidence function is at the
core of both $\hat{P}_{01}(s,t)$ and $\check{P}_{01}(s,t)$, and recent
research has investigated different proposals for estimating the
variance of an estimated cumulative incidence function \citep{Brau:Yuan:comp:2006,aa:cif:2010}.
Because of our representations~\eqref{eq:mm} and~\eqref{eq:check}, the
functional delta method justifies both use of the bootstrap and of a
normal limit. The tables report CIs both using the 25th and 75th
quantiles of the bootstrap estimates distribution and using a normal
approximation. Similar to the simulation study in
Section~\ref{sec:simu}, we find that $\hat{P}_{01}(s,t)$ and
$\check{P}_{01}(s,t)$ perform comparably.

{\color{black}Finally, Table~\ref{tab:hoho} displays the point
  estimates $\check{P}_{01}(s,t)$ together with the corresponding
  Aalen-Johansen estimates. Both estimators yield similar results.}

\begin{sidewaystable}
\begin{center}
\begin{tabular}{rrrrrrrrr}
\hline
& \multicolumn{4}{c}{New estimator} & \multicolumn{4}{c}{Meira-Machado estimator} \\
\hline
\multicolumn{1}{c}{$t$}&\multicolumn{1}{c}{$P_{01}(s, t)$}&\multicolumn{1}{c}{Variance}&\multicolumn{1}{c}{Bootstrap CI}&\multicolumn{1}{c}{Normal CI}&\multicolumn{1}{c}{$P_{01}(s, t)$}&\multicolumn{1}{c}{Variance}&\multicolumn{1}{c}{Bootstrap CI}&\multicolumn{1}{c}{Normal CI}\tabularnewline
\hline
$ 5$&$0.0234$&$1.95e-05$&[0.0152; 0.0323]&[0.0147; 0.032]&$0.0255$&$1.95e-05$&[0.0168; 0.0352]&[0.0162; 0.0347]\tabularnewline
$ 6$&$0.0314$&$2.45e-05$&[0.0219; 0.0413]&[0.0217; 0.0411]&$0.0342$&$2.45e-05$&[0.0244; 0.046]&[0.0236; 0.0448]\tabularnewline
$ 7$&$0.0363$&$2.82e-05$&[0.0258; 0.0469]&[0.0258; 0.0467]&$0.0395$&$2.82e-05$&[0.0286; 0.0517]&[0.0282; 0.0507]\tabularnewline
$ 8$&$0.0396$&$3.17e-05$&[0.0288; 0.051]&[0.0285; 0.0506]&$0.0431$&$3.17e-05$&[0.0315; 0.056]&[0.0313; 0.0549]\tabularnewline
$ 9$&$0.0452$&$3.57e-05$&[0.034; 0.0574]&[0.0335; 0.0569]&$0.0492$&$3.57e-05$&[0.0376; 0.0629]&[0.0366; 0.0618]\tabularnewline
$10$&$0.0476$&$3.76e-05$&[0.0361; 0.0596]&[0.0356; 0.0596]&$0.0518$&$3.76e-05$&[0.0392; 0.0655]&[0.0387; 0.0649]\tabularnewline
$11$&$0.0502$&$4.04e-05$&[0.0379; 0.0631]&[0.0377; 0.0627]&$0.0547$&$4.04e-05$&[0.0414; 0.0677]&[0.0414; 0.0679]\tabularnewline
$12$&$0.0512$&$4.04e-05$&[0.0388; 0.0637]&[0.0387; 0.0636]&$0.0557$&$4.04e-05$&[0.0432; 0.0695]&[0.0424; 0.0691]\tabularnewline
$13$&$0.0520$&$4.25e-05$&[0.0393; 0.0642]&[0.0392; 0.0648]&$0.0566$&$4.25e-05$&[0.0441; 0.0708]&[0.0432; 0.07]\tabularnewline
$14$&$0.0552$&$4.42e-05$&[0.0426; 0.068]&[0.0422; 0.0683]&$0.0601$&$4.42e-05$&[0.0471; 0.0747]&[0.0464; 0.0739]\tabularnewline
$15$&$0.0545$&$4.31e-05$&[0.0413; 0.0669]&[0.0416; 0.0673]&$0.0593$&$4.31e-05$&[0.0468; 0.0739]&[0.0456; 0.073]\tabularnewline
$20$&$0.0452$&$3.68e-05$&[0.0336; 0.0566]&[0.0333; 0.0571]&$0.0492$&$3.68e-05$&[0.037; 0.0632]&[0.0365; 0.062]\tabularnewline
$30$&$0.0258$&$2.09e-05$&[0.0174; 0.0346]&[0.0168; 0.0347]&$0.0280$&$2.09e-05$&[0.0191; 0.0391]&[0.018; 0.0381]\tabularnewline
$40$&$0.0176$&$1.60e-05$&[0.0101; 0.0256]&[0.0098; 0.0254]&$0.0192$&$1.60e-05$&[0.0115; 0.028]&[0.0108; 0.0275]\tabularnewline
$50$&$0.0100$&$9.03e-06$&[0.0045; 0.0163]&[0.0042; 0.0159]&$0.0109$&$9.03e-06$&[0.0055; 0.0179]&[0.0046; 0.0173]\tabularnewline
\hline
\end{tabular}
\caption{Estimate of $P_{01}(s, t),\, s = 3$ using the new
  estimator and Meira-Machado estimator, along with bootstrap 95\% CIs and CIs
  based on normal approximation}\label{tab:s3}
\end{center}
\end{sidewaystable}

\begin{sidewaystable}
\begin{center}
\begin{tabular}{rrrrrrrrr}
\hline
& \multicolumn{4}{c}{New estimator} & \multicolumn{4}{c}{Meira-Machado estimator} \\
\hline
\multicolumn{1}{c}{$t$}&\multicolumn{1}{c}{$P_{01}(s, t)$}&\multicolumn{1}{c}{Variance}&\multicolumn{1}{c}{Bootstrap CI}&\multicolumn{1}{c}{Normal CI}&\multicolumn{1}{c}{$P_{01}(s, t)$}&\multicolumn{1}{c}{Variance}&\multicolumn{1}{c}{Bootstrap CI}&\multicolumn{1}{c}{Normal CI}\tabularnewline
\hline
$ 7$&$0.0167$&$1.60e-05$&[0.0091; 0.0243]&[0.0089; 0.0246]&$0.0190$&$1.60e-05$&[0.0108; 0.0281]&[0.0101; 0.0278]\tabularnewline
$ 8$&$0.0208$&$1.95e-05$&[0.0119; 0.0296]&[0.0121; 0.0294]&$0.0236$&$1.95e-05$&[0.0143; 0.0336]&[0.0137; 0.0334]\tabularnewline
$ 9$&$0.0286$&$2.62e-05$&[0.0187; 0.0384]&[0.0186; 0.0386]&$0.0324$&$2.62e-05$&[0.0217; 0.0443]&[0.021; 0.0438]\tabularnewline
$10$&$0.0325$&$2.99e-05$&[0.0213; 0.0435]&[0.0218; 0.0432]&$0.0369$&$2.99e-05$&[0.0255; 0.0498]&[0.0248; 0.049]\tabularnewline
$11$&$0.0357$&$3.22e-05$&[0.0241; 0.0471]&[0.0246; 0.0468]&$0.0405$&$3.22e-05$&[0.0281; 0.0535]&[0.0279; 0.0531]\tabularnewline
$12$&$0.0379$&$3.41e-05$&[0.0262; 0.0494]&[0.0264; 0.0493]&$0.0430$&$3.41e-05$&[0.0308; 0.0559]&[0.0301; 0.0558]\tabularnewline
$13$&$0.0398$&$3.56e-05$&[0.0278; 0.0512]&[0.0281; 0.0515]&$0.0452$&$3.56e-05$&[0.0325; 0.0582]&[0.0321; 0.0583]\tabularnewline
$14$&$0.0438$&$4.10e-05$&[0.0309; 0.056]&[0.0312; 0.0563]&$0.0497$&$4.10e-05$&[0.036; 0.0644]&[0.0361; 0.0633]\tabularnewline
$15$&$0.0438$&$4.17e-05$&[0.0307; 0.0561]&[0.0311; 0.0565]&$0.0497$&$4.17e-05$&[0.0363; 0.0633]&[0.0359; 0.0635]\tabularnewline
$20$&$0.0402$&$4.01e-05$&[0.0277; 0.0529]&[0.0278; 0.0526]&$0.0456$&$4.01e-05$&[0.0324; 0.0593]&[0.032; 0.0593]\tabularnewline
$30$&$0.0233$&$2.43e-05$&[0.0139; 0.0336]&[0.0136; 0.0329]&$0.0264$&$2.43e-05$&[0.0157; 0.0374]&[0.0153; 0.0374]\tabularnewline
$40$&$0.0174$&$1.88e-05$&[0.0096; 0.0264]&[0.0088; 0.0259]&$0.0196$&$1.88e-05$&[0.0109; 0.0304]&[0.0101; 0.0292]\tabularnewline
$50$&$0.0102$&$1.16e-05$&[0.0042; 0.0175]&[0.0035; 0.0168]&$0.0115$&$1.16e-05$&[0.0049; 0.0195]&[0.0042; 0.0187]\tabularnewline
\end{tabular}
\end{center}
\caption{Estimate of $P_{01}(s, t)$, $s = 5$ using the new estimator and Meira-Machado estimator, along with bootstrap 95\% CIs and CIs based on normal approximation}\label{tab:s5}
\end{sidewaystable}

\begin{sidewaystable}
\begin{center}
\begin{tabular}{rrrrrrrrr}
\hline
& \multicolumn{4}{c}{New estimator} & \multicolumn{4}{c}{Meira-Machado estimator} \\
\hline
\multicolumn{1}{c}{$t$}&\multicolumn{1}{c}{$P_{01}(s, t)$}&\multicolumn{1}{c}{Variance}&\multicolumn{1}{c}{Bootstrap CI}&\multicolumn{1}{c}{Normal CI}&\multicolumn{1}{c}{$P_{01}(s, t)$}&\multicolumn{1}{c}{Variance}&\multicolumn{1}{c}{Bootstrap CI}&\multicolumn{1}{c}{Normal CI}\tabularnewline
\hline
$ 9$&$0.0165$&$2.02e-05$&[0.0087; 0.0266]&[0.0077; 0.0253]&$0.0192$&$2.02e-05$&[0.01; 0.0304]&[0.0087; 0.0297]\tabularnewline
$10$&$0.0215$&$2.63e-05$&[0.0119; 0.0329]&[0.0115; 0.0316]&$0.0251$&$2.63e-05$&[0.0139; 0.0381]&[0.013; 0.0371]\tabularnewline
$11$&$0.0269$&$3.33e-05$&[0.0167; 0.0398]&[0.0156; 0.0382]&$0.0313$&$3.33e-05$&[0.0186; 0.0459]&[0.0178; 0.0447]\tabularnewline
$12$&$0.0297$&$3.61e-05$&[0.0195; 0.0438]&[0.0179; 0.0414]&$0.0345$&$3.61e-05$&[0.0218; 0.0494]&[0.0206; 0.0484]\tabularnewline
$13$&$0.0334$&$4.16e-05$&[0.0218; 0.0478]&[0.0208; 0.0461]&$0.0389$&$4.16e-05$&[0.0248; 0.0546]&[0.024; 0.0538]\tabularnewline
$14$&$0.0385$&$4.92e-05$&[0.0257; 0.0546]&[0.0248; 0.0523]&$0.0448$&$4.92e-05$&[0.0301; 0.0617]&[0.0293; 0.0604]\tabularnewline
$15$&$0.0398$&$5.12e-05$&[0.0267; 0.0554]&[0.0258; 0.0538]&$0.0463$&$5.12e-05$&[0.0311; 0.0625]&[0.0309; 0.0617]\tabularnewline
$20$&$0.0364$&$4.70e-05$&[0.0229; 0.0514]&[0.0229; 0.0498]&$0.0424$&$4.70e-05$&[0.0288; 0.058]&[0.0275; 0.0573]\tabularnewline
$30$&$0.0245$&$3.28e-05$&[0.0139; 0.0375]&[0.0133; 0.0358]&$0.0287$&$3.28e-05$&[0.0166; 0.0424]&[0.0161; 0.0413]\tabularnewline
$40$&$0.0209$&$2.72e-05$&[0.0111; 0.0321]&[0.0107; 0.0311]&$0.0244$&$2.72e-05$&[0.0135; 0.0383]&[0.0121; 0.0367]\tabularnewline
$50$&$0.0130$&$1.77e-05$&[0.0057; 0.0222]&[0.0048; 0.0212]&$0.0152$&$1.77e-05$&[0.0061; 0.0259]&[0.0053; 0.025]\tabularnewline
\hline
\end{tabular}
\end{center}
\caption{Estimate of $P_{01}(s, t)$, $s = 7$ using the new estimator and Meira-Machado estimator, along with bootstrap 95\% CIs and CIs based on normal approximation}\label{tab:s7}
\end{sidewaystable}

\begin{table}[htb]\color{black}
  \centering
  \begin{tabular}{r|rr|rr|rr}
    \multicolumn{1}{c|}{$t$}&\multicolumn{1}{c}{$\check P_{01}(3, t)$} & Aalen- & \multicolumn{1}{c}{$\check P_{01}(5, t)$} & Aalen-& \multicolumn{1}{c}{$\check P_{01}(7, t)$} & Aalen-\tabularnewline
    {} & {} & Johansen & {} & Johansen& {} & Johansen\\
\hline
$ 5$&$0.0234$&$0.0266$& {}     & {}     & {}     &\tabularnewline
$ 6$&$0.0314$&$0.0359$& {}     & {}     & {}     &\tabularnewline
$ 7$&$0.0363$&$0.0411$&$0.0167$&$0.0200$& {}     &\tabularnewline
$ 8$&$0.0396$&$0.0446$&$0.0208$&$0.0250$& {}     &\tabularnewline
$ 9$&$0.0452$&$0.0515$&$0.0286$&$0.0343$&$0.0165$&$0.01987$\tabularnewline
$10$&$0.0476$&$0.0533$&$0.0325$&$0.0376$&$0.0215$&$0.02498$\tabularnewline
$11$&$0.0502$&$0.0559$&$0.0357$&$0.0419$&$0.0269$&$0.03141$\tabularnewline
$12$&$0.0512$&$0.0569$&$0.0379$&$0.0440$&$0.0297$&$0.03481$\tabularnewline
$13$&$0.0520$&$0.0578$&$0.0398$&$0.0460$&$0.0334$&$0.03813$\tabularnewline
$14$&$0.0552$&$0.0612$&$0.0438$&$0.0503$&$0.0385$&$0.04389$\tabularnewline
$15$&$0.0545$&$0.0605$&$0.0438$&$0.0505$&$0.0398$&$0.04503$\tabularnewline
$20$&$0.0452$&$0.0509$&$0.0402$&$0.0445$&$0.0364$&$0.04218$\tabularnewline
$30$&$0.0258$&$0.0292$&$0.0233$&$0.0270$&$0.0245$&$0.02726$\tabularnewline
$40$&$0.0176$&$0.0204$&$0.0174$&$0.0196$&$0.0209$&$0.02061$\tabularnewline
$50$&$0.0100$&$0.0115$&$0.0102$&$0.0111$&$0.0130$&$0.01165$\tabularnewline
  \end{tabular}
  \caption{Point estimates $\check P_{01}(s, t)$ as in Tables~\ref{tab:s3}--\ref{tab:s7} and corresponding Aalen-Johansen estimates.}
  \label{tab:hoho}
\end{table}

\clearpage

\section{Discussion}\label{sec:disc}
We have demonstrated how to use competing risks techniques for
estimating transition probabilities in a non-Markov illness-death
model without recovery. For ease of presentation, we have focused on
estimating $P_{01}(s,t)$. Our first estimator, $\hat{P}_{01}(s,t)$
from~\eqref{eq:mm}, is algebraically equal to the estimator of
\cite{Meir:Ua:Cada:nonp:2006} who derived it using Kaplan-Meier
integrals. We have also given a new IPCW representation of the
estimator, which we have then used to find a computationally simpler
estimator, $\check{P}_{01}(s,t)$ from~\eqref{eq:check}.

To discuss the intrinsic properties of the proposed estimators, it is
useful to consider the special case where the process is fully
observed for all cases (uncensored data). Then, transition
probabilities can be consistently estimated by ratios of crude counts
also when the process is non-Markov. In fact, for uncensored data
both~$\hat{P}_{01}(s,t)$ and~$\check{P}_{01}(s,t)$ reduce to
\begin{equation}
\frac{\sum_{i=1}^n \eins\{X_s^{(i)}=0, X_t^{(i)}=1\}}{\sum_{i=1}^n \eins\{X_s^{(i)}=0\}},\label{eq:1}
\end{equation}
where the {\color{black}superscript}~$(i)$ indicates the $i$th
replicate of~$n$ i.i.d.\ copies of the multistate process.
This is in analogy to many estimators of the state
  occupation probabilities which reduce to the usual multinomial
  estimators for complete data. In~\eqref{eq:1}, each individual
contributes with equal weight~$1/n$ to the sum in the nominator and in
the denominator.

For right-censored data, the status of the process is unknown after
the individual end of study time. From an IPCW perspective, the idea
underlying~$\hat{P}_{01}(s,t)$ is to restrict the summation in
\eqref{eq:1} to the individuals not lost to follow-up before time $t$
and to re-weight their contributions by the probability of not being
lost to follow-up. The weights are based on a Kaplan-Meier estimate of
the censoring distribution using the censored observations of~$T$,
see~\eqref{eq:ipcw}.

However, some individuals will be lost to follow-up in the initial
state and others in the disease state. This information is not used
by~$\hat{P}_{01}(s,t)$, but~$\check{P}_{01}(s,t)$ uses such
information, see~\eqref{eq:TC}. Theoretically, $\check{P}_{01}(s,t)$
is therefore more efficient, but the simulation results and the
practical data example found comparable performance. The practical
advantage of $\check{P}_{01}(s,t)$ is that it is computationally
simpler.

{\color{black}A further advantage of $\check{P}_{01}(s,t)$ is that,
  being an Aalen-Johansen estimator of the limit of a certain
  cumulative incidence function, it gives direct access to competing
  risks methodology, as we have demonstrated by also allowing for
  left-truncated data.} In the context of hospital-acquired
infections, such a delayed study entry may arise if patients are not
followed since admission but conditional on detection of an infectious
organism such as Methicillin-Resistant Staphylococcus Aureus as in
\citet{de2011multistate}.

So far, a drawback of the estimation procedures as outlined both in
the present paper and in \cite{Meir:Ua:Cada:nonp:2006} is that we
require the support of the distribution of~$T$ to be contained in the
support of the distribution of~$C$ in order to be able to estimate the
limit of a cumulative incidence function, see~\eqref{eq:limit}. This
is not a restriction for our motivating data situation, but the
assumption is often not fulfilled in other medical applications. The
problem can be circumvented by `artificial censoring' black
  as, e.g., in \citet{quale06:_local}.

To be specific, consider the fixed, but arbitrary time pair $s\le t$
and assume that $s,t < \inf\{v:\P(C>v)=0\}$. Then there is a $\tau>t$
with $P(C>\tau)>0$. The idea is to consider the modified random
variables~$(\min(T_0, \tau), \min(T, \tau))$ instead of $(T_0,T)$.
Their distributions coincide on $[0,\tau)\times [0,\tau)$, which
includes the bivariate time point of interest $(s,t)$, and $\min(T,
\tau)$ is less than $\inf\{v:\P(C>v)=0\}$ by construction. We can then
use the estimation techniques as outlined earlier, but using the
modified data. Note that the data do change. E.g., if~observation of
$T$ is censored after the chosen~$\tau$, the modified variable~$\min(T,
\tau)$ has been observed.

{\color{black}Finally, our limited simulation study indicated that the
  Aalen-Johansen estimator may competitively estimate transition
  probabilities in small samples even in the absence of the Markov
  property. This is not unlike the findings of
  \cite{gunnes2007estimating} for estimating state occupation
  probabilities.
}

{\color{black}
\section*{Appendix}
The aim of the appendix is to show that our initial estimation
procedure based on the competing risks process~$\kappa$ is
algebraically identical with the proposal of
\cite{Meir:Ua:Cada:nonp:2006}. The idea of their estimator is to
consider~$T_0$ as a covariate for the event time~$T$ and to use
Stute's estimator for a Kaplan-Meier integral with a covariate
\citep{stute93:_consis}.

For the purpose of comparison, note that the formulation of
Meira-Machado et al.\ is based on latent transition times between the
states of the illness-death model. These authors then consider
censored variants of such latent times, provided they are observable.
Meira-Machado et al.\ then arrive at censored variants of $(T_0,T)$,
which will be our starting point. Also note that because $T_0$ will be
considered as a covariate for a Kaplan-Meier integral with respect
to~$T$, we will only need an event indicator for the latter. This will
further simplify the notation. We will also use that $T_0$ has been
observed, if $T$ has been observed, because $T_0\le T$.

Stute's method requires that the parameter of interest can be
formulated as an integral with respect to the joint distribution of
$(T_0,T)$,
\begin{displaymath}
  \int\phi(z,y)\,\P^{T_0,T}(\dif z,\dif y).
\end{displaymath}
Again focussing on $P_{01}(s,t)$ for ease of presentation, the
Meira-Machado et al.\ estimator relies on estimating the above display
for $\phi(z,y)=\eins(s<z\le t, t<y)$.

Assume~$n$ i.i.d.\ data~$(\tilde T_{0i}, \tilde T_i, \xi_i)$,
$i=1,\ldots n$, where the tilde indicates a censored observation,
e.g., $\tilde T_i =\min(T_i, C_i)$, $\xi_i$ is the event
indicator~$\eins(T_i\le C_i)$, and the index~$i$ indicates the $i$th
individual. Stute's method (and the estimator of Meira-Machado et al.)
is based on the ordered data $T_{(1)}\le \ldots \le T_{(n)}$ with
$(\xi_{[i]}, T_{0[i]})$ attached to $T_{(i)}$. Again for ease of
presentation, we assume no ties in the data; \cite{stute93:_consis}
discusses how to arbitrarily break ties if present. Note that our
formulation of the estimators does allow for ties.

The Meira-Machado et al.\ estimator of~$P(s<T_0\le t, t<T)$ is
\begin{displaymath}
    \sum_{i=1}^n
  \prod_{j=1}^{i-1}\left(1-\frac{\xi_{[j]}}{n-j+1}\right)\frac{\xi_{[i]}}{n-i+1} \phi(T_{0[i]}, T_{(i)}).
\end{displaymath}
Using the counting process notation introduced earlier, the above
display equals
\begin{displaymath}
    \sum_{i=1}^n
  \prod_{j=1}^{i-1}\left(1-\frac{\Delta N(\tilde T_{(j)})}{Y(\tilde T_{(j)})}\right) \frac{\Delta N(\tilde T_{(i)})}{Y(\tilde T_{(i)})}\phi(T_{0[i]}, T_{(i)}).
\end{displaymath}
We note two things about the last display: Firstly, because the sum
runs over all individuals and because addition and multiplication are
each commutative, ordering is not needed. Secondly, if $\Delta
N(\tilde T_{i})=1$, then $\tilde T_{i}= T_i$ and $\tilde T_{0i}=
T_{0i}$. Hence, we have $\Delta N(\tilde T_{i}) \cdot \phi(T_{0i},
T_{i}) = \Delta N_1(\tilde T_{i})$. As a consequence, the
Meira-Machado et al.\ estimator of~$P(s<T_0\le t, t<T)$ equals our
competing risks-type estimator~\eqref{eq:est1} and hence our
estimator~\eqref{eq:mm} equals their estimator of~$P_{01}(s,t)$.  }

\end{document}